\input  phyzzx
\input epsf
\overfullrule=0pt
\hsize=6.5truein
\vsize=9.0truein
\voffset=-0.1truein
\hoffset=-0.1truein

%
%
\def\half{{1\over 2}}
\def\IC{{\ \hbox{{\rm I}\kern-.6em\hbox{\bf C}}}}
\def\IR{{\hbox{{\rm I}\kern-.2em\hbox{\rm R}}}}
\def\IZ{{\hbox{{\rm Z}\kern-.4em\hbox{\rm Z}}}}

\def\sIR{{\hbox{{\sevenrm I}\kern-.2em\hbox{\sevenrm R}}}}

\def\sym{super Yang-Mills theory}
\def\til{\tilde g^2}
\def\mt{M(atrix) Theory}
\def\tild{$\tilde D$0-brane}
%
%
\hyphenation{Min-kow-ski}

\rightline{SLAC-PUB-7431}
\rightline{SU-ITP-97-09}
\rightline{UTTG-10-97}
\rightline{March 1997}
\rightline{hep-th/9703102}

\vfill

%
%
\title{The Incredible Shrinking Torus\foot{Submitted to Nuclear Physics B.}}

\vfill

%
%
\author{W.Fischler$^1$
\foot{Supported in part by
NSF Grant PHY-9511632.}, E. Halyo,   A.
Rajaraman$^3$\foot{Supported in part by
the Department of Energy
under contract no. DE-AC03-76SF00515.} and
L. Susskind$^2$\foot{Supported
by NSF grant PHY-9219345.}}

\vfill

\address{$^1$Theory Group,Department of Physics,\break University of Texas,
Austin,TX 78712}
\address{$^2$Department of Physics,  Stanford University\break Stanford, CA
94305-4060}
\address{$^3$Stanford Linear Accelerator Center,  Stanford University\break
Stanford, CA 94305}

\vfill

%
%

Using M(atrix) Theory, the  dualities of toroidally compactified M-theory
can be formulated as properties of super Yang Mills theories in various
dimensions. We consider the cases of compactification on one, two, three,
four and five dimensional tori. The dualities required by string theory
lead to conjectures of remarkable symmetries and relations between field
theories as well as extremely unusual dynamical properties.  By studying
the theories in the limit of vanishingly small tori, a wealth of
information is obtained about strongly coupled fixed points of \sym \ in
various dimensions.  Perhaps the most striking behavior, as noted by
Rozali in this context, is the emergence of an additional dimension of space in
the case of
a four torus.
\vfill
\vfill\endpage

%
%
%
\REF\bfss{T. Banks,W. Fischler,S. Shenker and L. Susskind,hep-th/9610043.}
\REF\wati{W.Taylor,hep-th/9611042.}
\REF\sethi{S. Sethi and L. Susskind,hep-th/9702101.}
\REF\tdual{L. Susskind,hep-th/9611164.}
\REF\grt{O.J. Ganor, S. Ramgoolam and W.Taylor,hep-th/9611202.}
\REF\Seib{N. Seiberg, hep-th/9608111.}
\REF\Seibe{N. Seiberg, hep-th/9609161.}
\REF\motl{L. Motl,hep-th/9701025.}
\REF\IIAbs{T. Banks and N. Seiberg,hep-th/9702187.}
\REF\IIAVV{R. Dijkgraaf,E. Verlinde and H. Verlinde, hep-th/9703030.}
\REF\IIB{J. Schwarz, Phys. Lett.  {\bf B360} (1995) 13, hep-th/9508143; P.
Aspinwall, hep-th/9508154.}
\REF\bss{T.Banks, N.Seiberg and S.Shenker,hep-th/9612157.}
\REF\rozali{M. Rozali,hep-th/9702136.}
\REF\MS{J. Maldacena and L. Susskind, Nucl. Phys. {\bf B475} (1996) 670,
hep-th/9604042.}
\REF\DLCQ{H.C. Pauli and S.J. Brodsky,Phys.Rev.{\bf D32} (1985) 1993.}


%
%

%
%
\chapter{Introduction}

M(atrix) theory [\bfss] is a nonperturbative formulation of string theory in
terms
of certain supersymmetric Yang-Mills theories. The known properties of the
field theories can often illuminate the behaviour of string and M theory.
For example the non-renormalization theorems of super Yang-Mills theories
are essential
to an understanding of causality in space time and the existence of
asymptotic states. Another example is  the use of electric-magnetic duality
of four dimensional super Yang-Mills theory to establish T-duality
non-perturbatively in
the case of the  three torus [\wati,\sethi,\tdual,\grt]. The other side of the
coin is the use of
known or expected properties of M- and string theory to formulate
conjectures about field theory. In this paper we will be interested in the
behavior of super Yang-Mills theories,  with 16 real supersymmetries, that
can be derived by considering M(atrix) theory compactifications on small
tori. By small we will mean scales typically of order or smaller than the
11 dimensional Planck scale. In the following sections we
will consider the cases of tori of dimension 1, 2, 3, 4, 5 and 6.

We will see that by studying the limit of vanishing  size tori, we will
be driven to the limit of strongly coupled \sym.
This limit  is intrinsically interesting for the field theories,
especially  when the space-time dimensionality of the field theory
is not 4. In these cases the theories are either super-renormalizable or
non-renormalizable. In the former case the theory becomes strongly coupled
in the infrared while in the latter the strong coupling occurs in the
ultraviolet. In each case the difficult  region is described by a strongly
coupled fixed point which can have very nontrivial and often surprising
features.

Let us consider a  $D$ dimensional  \sym \ compactified on a
$ D-1$ torus with dimensions all of order $\Sigma$. The Yang-Mills
coupling constant, $g^2$ has dimension $m^{4-D}$.   The coupling constant
is unrenormalized due to the high degree of supersymmetry. However that
does not mean that the strength of interaction does not run with mass
scale.  If we introduce a scale
$\mu$, a dimensionless running coupling constant $\til$ may be defined.
$$
\til(\mu) = g^2 \mu^{D-4}
\eqn\oneone
$$
which measures the strength of interaction at length scale $\mu^{-1}$.  In
particular, the coupling strength at  scales of order  the torus size
$\Sigma$ is  defined to be $\til$
$$
\til=g^2 \Sigma^{4-D}
\eqn\onetwo
$$
The behavior of the theory  as the length scale varies can be studied by
fixing $g$ and letting the size of the torus vary. Thus we study the
ultraviolet (infrared) properties of the theory by  decreasing (increasing)
$\Sigma$. Alternately we may keep $\Sigma\ $ fixed and vary g.   For $D<4$,
going to strong dimensionless coupling probes the infrared behavior of
\sym. On the other hand, for $D>4$, strong coupling probes the ultraviolet
behavior. We shall see that in both cases we obtain information by studying
M(atrix) theory on vanishingly small tori.

Before stating our results we define some conventions.  M(atrix)
theory is defined by identifying the large $N$ limit of $U(N)$ \sym \ with
the infinite momentum limit in which the 11th direction is chosen as the
longitudinal direction. The theory is assumed to be compactified in this
direction on a circle of size $R$. Physical quantities all become
independent of $R$ in the large $N$ limit. We will consider compactifying
some subset of the transverse dimensions on a d-torus. This d-torus never
includes the 11th direction which may be assumed noncompact in the large
$N$ limit.

The various length scales that we encounter are the following. The 11
dimensional Planck scale of M-theory is defined to be
$l_{11}$, the  $d=D-1$ dimensional torus that M-theory is compactified on
has sides $L_1,L_2,....L_d$. Each length is assumed to  be of the same
order of magnitude $L$. The \sym
\ is also compactified on a dual torus whose dimensions are defined to be
$\Sigma_1, \Sigma_2, ...\Sigma_d$. Each of the $\Sigma_i$  is of the same
order of magnitude which is however, in general very different than the
original
length $L_i$. As
mentioned above, the Yang-Mills  coupling is
$g$ and the dimensionless coupling is
$\til$.

A few words about transversely compactified M(atrix) Theory
will help eliminate later confusion. In the original discussion of [\bfss]
the starting
point was the theory of  $D0$-branes in type IIA theory when the compact
direction is identified as the ``longitudinal" direction.  These objects
became the carriers of longitudinal momentum  ($p_{11}$) in an infinite
momentum or light cone
description. The number of these $D0$-branes is eventually allowed to go to
infinity.
In order to distinguish these carriers of $p_{11}$ from other $D0$-branes
which will
emerge when we compactify transverse dimensions, we will introduce the
notation \tild \ for the carriers of longitudinal momentum. Thus one would
say that the infinite momentum limit is obtained by letting the number of
\tild s tend to infinity.

The results of our analysis are as follows.

\item {1.} For the case of the 1-torus, shrinking  yields a
theory which is equivalent, through the M(atrix) connection, to
uncompactified  perturbative type IIA string theory.
\itemitem{a)} The spectrum of the
theory is in one to one correspondence with the spectrum of perturbative
string states. The limit $L \to 0$  is described by a stongly
coupled \sym \ [\wati] with dimensionless coupling $\til =(2\pi)^4{l_{11}^3
\over L^3}$.
On the string side it describes IIA string theory with string coupling
$g_s^2 = {2\pi\over \til}$. Thus the limit $L \to 0$ is free string theory.
The \sym \  has a strong coupling expansion which is just the usual string
loop expansion.
\itemitem{b)}The $U(N)$ \sym \ has a normalizable ground state
with unbroken $U(N)$ symmetry and an excited state spectrum with a mass
gap equal to ${2\pi \over N \Sigma}$. The theory also has a variety of Higgsed
vacua with symmetry broken to products of smaller
$U(N_i)$ factors. The normalizability of the ground state has implications
for the behavior of D1-branes in IIB theory. It requires the existence of
normalizable bound states for a collection of $N$ D1-branes.
\itemitem{c)}We identify the states of the \sym \ which describe the
$D0$-branes
of the transverse compactification.

\item{2.} For compactification on a 2-torus, the vanishing torus limit  is 10
dimensional type IIB theory [\IIB]. More generally, compactification on a
2-torus
leads to type IIB string theory with 1 compact space dimension. The infinite
momentum description has a manifest $O(7)$ rotation invariance. The coupling
constant is of order 1 unless the sides of the torus are very different in
which case it becomes weakly coupled. Shrinking the torus decompactifies the
compact dimension and leads to an enhanced rotational symmetry $O(8)$. This
implies that the 2+1 dimensional \sym \ also develops the enhanced $O(8)$
symmetry at strong coupling. Since in this case, it is the infrared behavior
which is governed by strong coupling, the implication is that the I.R.
behavior is
$O(8)$ invariant [\sethi,\IIAbs].

\item{3.}
The shrunken 3-torus limit leads back to decompactified M-theory. This can
be seen from T-duality. On the  \sym \ side, the symmetry under T-duality
becomes symmetry under inverting the coupling constant  $g \to {2 \pi
\over g}$. This is just the Montonen Olive electric-magnetic duality of
four dimensional \sym. Had it not already been known, this would have
been the prediction following from compactification on a 3-torus. In this
case, the theory is scale invariant.

\item{4.} The four torus is perhaps the most interesting example of all. It
is described by 4+1 dimensional \sym \ [\Seib].
\itemitem{a)}Shrinking the 4-torus to zero leads to weakly coupled string
theory in 7 noncompact space-time dimensions. The coupling vanishes in
the limit and the compactification radii are all of string scale.
\itemitem{b)}The U-duality group of the theory is $SL(5;Z)$. This is the
symmetry group of the 5-torus. Following Rozali [\rozali], we find that the
\sym \
develops a new 5th spatial direction whose radius grows as the M-theory
torus shrinks. The duality group is a reflection of the 5-dimensional
behavior.
\itemitem{b)} The origin of the new direction is light states which
come down  as the Yang-Mills coupling increases. These light states carry
instanton charge which may be identified as the momentum in the 5th
spatial direction.
\itemitem{c)}In this case  strong coupling behavior governs the
ultraviolet behavior of the theory. This means that  the short
distance behavior of 4+1 dimensional \sym \ is controlled  by a fixed point
 describing a five dimensional field theory with one very large
dimension!

\item{5.}Compactification on the 5-torus [\Seibe] exhibits a phenomenon which
in a
sense is the reverse of what happens in the 4-torus case. The strongly
coupled limit of this theory is described by a 4+1 dimensional system,
namely \sym \ with a coupling of order unity.  On the M-theory side the
limiting theory can be dualized to M-theory on a 4-torus whose size is of
order the 11 dimensional Planck scale. It can also be described as IIA
string theory at intermediate coupling, compactified on a torus of string
scale size.

\item{6.} For d equal to or greater than 6 we find that we are unable to
dualize the theory into any form in which it can be reliably analyzed. Any
attempt to dualize it to weak or intermediate coupling produces
ultra-small compactification radii. Applying T-duality to increase the
compactification radii inevitably leads to ultra-large coupling. From the
field-theoretic side this may be connected with the lack of superconformal
fixed points in space-time dimension greater than 6. What it means on the
string or M-theoretic side is unclear. Perhaps it signals some sort of
non-perturbative breakdown of toroidal compactification. Resolving this
point could have important implications for 4-dimensions.

%
\chapter{The 1-torus}

Compactification of M(atrix) theory on a circle of circumference $L$
leads to a
$1+1$ dimensional $U(N)$ \sym \ containing a vector field, eight
scalars and
their fermionic partners [\wati]. We will take the compact transverse direction
of
the
M(atrix)  theory to be the 9th direction. The Lagrangian is
$$
L={1\over g^2}\int^{\Sigma} d\sigma Tr\left(-{1\over 4} F_{\mu \nu}^2 +
D_{\mu}\phi_i^2 +[\phi_i,\phi_j]^2 +fermions \right)
\eqn\twoone
$$
where $\phi_i$ represents the 8 adjoint scalars, $D_{\mu}$ is the covariant
derivative, $\Sigma$ is the circumference of the circle parametrized by
$\sigma$.

To determine the relation between $L$ and the field theoretic
parameters $g^2,
\Sigma$ we use the method of [\tdual] which entails comparing certain
energy scales. Suppose that the $U(N)$ symmetry is broken to $U(N-1)
\times U(1)$ with the $U(1)$ factor describing the motion of a single
\tild.  In the temporal gauge
$A_0=0$ the Lagrangian for the homogeneous  mode of the $U(1)$
gauge field is
$$
{1\over{2 g^2}}\Sigma \dot A_1^2
\eqn\twotwo
$$
Furthermore $A\Sigma$ is an angle whose momentum is quantized in
integers. Thus
the energy stored in the electic field has the form
${g^2\over{2 }}\Sigma n^2$
with $n$ being an integer. This energy is to be identified with the kinetic
energy of a  $\tilde D$0-brane which has transverse momentum
$n$ in the 9th direction. This energy is of the form ${{p_9}^2\over2 p_{11}}$
which is
equal to $(2\pi)^2n^2 R \over2L^2$. Thus we find
$$
g^2 \Sigma = {(2\pi)^2R\over L^2}
\eqn\twothree
$$

Next we equate the energy of a membrane wrapped around $R$ and $L$ to the
energy of a single quantum in the field theory. The membrane energy is
${RL\over 2\pi l_{11}^3}$ and the energy of the quantum is $2\pi\over \Sigma$.
This
gives
$$
{2\pi\over \Sigma} ={RL\over 2\pi l_{11}^3}
\eqn\twofour
$$
The parameter $\Sigma$ is the size of the circle upon which the super
Yang-Mills field theory is compactified. We will refer to it as the field
theory
compactification radius. From \twofour \ we see that $\Sigma$ becomes large
when $L$ becomes small.

 Equations
\twothree
\  and
\twofour
\ determine the parameters
$\Sigma$  and
$g^2$. From these a single dimensionless parameter $\tilde g^2$ can be obtained
$$
\til =g^2\Sigma^2=(2\pi)^4{l_{11}^3\over L^3  }
\eqn\twofive
$$
The parameter $\til$ is a measure of the strength of interaction at length
scales of order $\Sigma$, that is the largest scale in the theory. In what
follows we will be especially interested in the limit in which $L$ is much
smaller than $l_{11}$. This is the limit in which the dimensionless
coupling
$\til$ is very large. We know very little about this theory but if we
believe
the M(atrix) theory connection then we can determine a great deal by
connecting it with weakly coupled type IIA string theory. Before we
do so it
is worth pointing out that this same field theory governs the behavior of a
collection of $N$ D1-branes wound around a large cycle in type IIB string
theory.

The connection with type IIA theory comes from the fact that M-Theory
compactified on a circle of circumference $L$ becomes weakly
coupled IIA string
theory as $L$ tends to zero. The precise connection is that
the string length and string coupling $g_s$ are given by
$$
{l_s^2\over 2}=\alpha'={2\pi l_{11}^3 \over L }
\eqn\twosix
$$
and
$$
g_s^2 = {1\over (2\pi)^3}{L^3 \over l_{11}^3}
\eqn\twoseven
$$
Evidently the dimensionless Yang-Mills coupling $\til$ and the IIA string
coupling are inversely related to one another.
$$
g_s^2={2\pi\over \til}     
\eqn\twoeight           
$$                      
Thus the limit of strongly coupled super Yang-Mills theory is equivalent to
free string
theory. As an example, consider the properties of the ground state.
It is not
known whether the field theory has a normalizable ground state or if
the
ground state wave function spreads out along the flat directions of
the moduli
space.  To answer this we note that the ground state of string theory
in the
infinite momentum frame with longitudinal momentum $R/N$ is the graviton
whose
energy is $Rp^2 \over 2N$ where $p$ is the transverse momentum in the
uncompactified directions. For
$p=0$ the energy is minimized. This state, being an eigenstate of
transverse
momentum is of course not normalizable but this center of mass factor is
associated with the
$U(1)$ factor of the gauge group. This can be trivially factored off
so that we
are really considering the normalizability of the $SU(N)$ theory. Now
apart from
its center of mass motion a single graviton should be a normalizable
state.
Thus it follows that the $SU(N)$ 1+1 dimensional super Yang-Mills
theory on a
circle has a normalizable ground state. This fact has implications for the
theory of D1-branes in type IIB theory. It says that a collection of  $N$
D1-branes wound around a compact cycle has a single normalizable
bound state.
Note that this result is in direct conflict with claims in the literature
{}.

Free string theory has superselection sectors corresponding to
states with any
number of disconnected strings. The corresponding phenomenon in
super Yang-Mills theory is
the Higgs phenomenon in which $U(N)$ breaks down into
$U(N_1)\times U(N_2)\times U(N_3)...$ [\motl,\IIAbs,\IIAVV] The individual
factors
describe the
separated strings. Let us concentrate on the unbroken sector
of the \sym \ and
consider the excited state spectrum. This corresponds to the
spectrum of
excited free strings. According to free string theory in the infinite
momentum frame the energy of the first excitation is
$$
E={M^2 \over 2P_{11}} =M^2 { R \over 2N}
\eqn\twonine
$$
where $M^2 ={2 \over \alpha'}$. This translates to an energy gap in the \sym
 .
$$
E={2\pi \over N \Sigma}
\eqn\twoten
$$
This gap is surprisingly small. The natural gap for a massless field theory
defined on a circle of size $\Sigma$ would be $2\pi \over \Sigma$. The gap in
\twoten \ is $N$ times smaller.This fact has  also been noted by Motl and by
Banks and Seiberg and is believed to be related to the phenomenon
reported in [\MS].

The string theory connection indicates that there should exist a strong
coupling expansion for the field theory. Since the  super Yang-Mills coupling
is
inverse to
the string coupling, the field theoretic strong coupling expansion is
just the
familiar genus expansion of 10 dimensional IIA theory.

Consider next the  states which are excited by having a single unit of
momentum along the compactified 9th direction. The momentum in the
compactified direction is proportional to the $U(1)$ electric field (Note
that in this case we are not discussing the $U(1)$ factor which occurs when
$U(N) \to U(N-1) \times U(1)$ but rather the $U(1)$ which commutes with
 $SU(N)$). A
single quantum of electric field carries energy
$$
E_E ={g^2 \Sigma \over 2N}=
{R\over 2N}\left({2\pi\over L}\right)^2
\eqn\twoeleven
$$
This field theoretic energy is interpreted as an energy in the infinite
momentum frame of M-theory. It corresponds to a mass
$$
M_E={2\pi\over L}={1\over g_s\sqrt{\alpha'}}
\eqn\twotwelve
$$
This is exactly the mass of a $D0$-brane of the weakly coupled
type IIA string
theory. We emphasize that this object is not the original \tild \
that carries
longitudinal momentum  and whose number goes to infinity in the infinite
momentum limit. These new objects are the massive $D0$-branes of the
transversely compactified theory whose existence is required by weakly
coupled
string theory. Their presence in the spectrum of the \sym \ constitutes a
consistency check for M(atrix) theory. It would be very interesting to study
the sector of the theory with $n$ units of electric energy corresponding to
$n  D0$-branes. Using purely field theoretic methods it should be possible to
discover that this sector is described by $U(n)$ \sym \ in 0+1 dimensions.
%
%
\chapter{The 2-torus}

Next consider \mt \ compactified on a small 2-torus. The length of the
sides of the
torus are $L_1,L_2$. The appropriate field theory is 2+1 dimensional \sym \
 on a
2-torus with sides $\Sigma_1, \Sigma_2$ and coupling constant $g$. Following
 the
procedure in the previous section [\tdual] we find that these parameters
are
connected by the following equations.
$$
g^2=(2\pi)^2{R \over L_1L_2}
\eqn\threeone
$$
$$
\Sigma_i=(2\pi)^2{l_{11}^3 \over L_iR}
\eqn\threetwo
$$
where $i=1,2$. A dimensionless coupling constant $\til$  can also be
 defined by
multiplying
$g$ by the appropriate power of the area $\Sigma_1\Sigma_2$.
$$
\til =g^2 (\Sigma_1\Sigma_2)^{\half}=(2\pi)^4{l_{11}^3 \over (L_1L_2)^{3\over
2}
}
\eqn\threethree
$$
Again, when the volume $L_1L_2$ becomes small, the dimensionless Yang-Mills
coupling becomes large. The new feature in this case is the existence of
magnetic flux through
the $\Sigma_1\Sigma_2$ torus. The abelian magnetic flux is quantized in
integer multiples of $2\pi$.
One easily finds that the energy associated with the flux is given in terms
of Yang-Mills quantities by
$$
E_M ={(2\pi)^2 \over 2Ng^2\Sigma_1 \Sigma_2}
\eqn\threefour
$$
It can be reexpressed in terms of the $L_i$, $R$ and $l_{11}$.
$$
E_M = {R\over 2N}{L_1^2 L_2^2 \over (2\pi)^4l_{11}^6}={1\over 2p_{11}}{L_1^
2
L_2^2 \over
(2\pi)^4l_{11}^6}
\eqn\threefive
$$
The mass associated with this energy is
$$
M_M={L_1L_2\over (2\pi)^2l_{11}^3}
\eqn\threesix
$$

The expression in eq. \threesix \ is precisely the mass of a two brane in
M-theory [\grt],
wrapped on the 2-torus. When the 2-torus shrinks to zero size, the magnetic
energy
gives rise to a dense spectrum of low energy states. In all the higher
dimensional examples we will consider, a similar  infinity of light states
come down as the
torus shrinks. The interpretation of these states is one of the main themes
of this
paper. Surprisingly, the interpretation is quite different in each dimension.

In the present case the interpretation is well known [\IIB,\sethi]. A new
dimension, $Y$, opens
up
and becomes decompactified as $L_i \to 0$. The quantized flux becomes the
Kaluza-Klein momentum in the new direction. To compute the circumference
$L_y$ of
the new direction we equate the energy in eq. \threesix \ to the energy of
the first
Kaluza-Klein excited state. This gives
$$
L_y=(2\pi)^3{l_{11}^3 \over L_1L_2}
$$
or more symmetrically
$$
L_yL_1L_2=(2\pi)^3l_{11}^3
\eqn\threeseven
$$

Let us next consider the energy stored in the $U(1)$ electric field.
Following the
same logic as in eqs. \twoeleven \ and \twotwelve \ we find that the energy
 in a
single quantum of electric field in the $1$ direction   is
$$
E_E ={g^2 \over 2N}{\Sigma_1 \over \Sigma_2} ={R \over
2N}\left({L_yL_2\over (2\pi)^2l_{11}^3}\right)^2
\eqn\threeeight
$$
The mass of the state is
$$
M_E={L_yL_2\over (2\pi)^2l_{11}^3}
\eqn\threenine
$$
This mass corresponds to the energy of a membrane wrapped around the directions
$(2,Y)$. As  $L_2$ tends to zero, the configuration becomes a string, wound
around
the $Y$ direction. In fact it is a type IIB string. The coupling constant of
 the
type IIB string theory is given by
$$
g_s={L_2 \over L_1}
\eqn\threeten
$$
Thus we see that the electric field in the $1$ direction corresponds to the
winding
number of type IIB strings. If the ratio ${L_2\over L_1}$ is small these
strings
are weakly coupled. We can also consider the electric field in the 2 direction
which by an identical argument must also correspond to some kind of string.
These
strings are easily identified as the D-strings of type IIB theory. The
$SL(2;Z)$
symmetry of the IIB theory is just the geometric symmetry of the shrinking two
torus. All of these facts lead to the conclusion that in the large N limit,
2+1
dimensional \sym \ is equivalent to type IIB string theory.

The consequences of this conclusion for the \sym \ include many predictions
about
the spectrum of the theory. For example the energy gap in the un-Higgsed
sector must
be of the form
$$
E={1\over 2N}{1 \over(\Sigma_1 \Sigma_2)^{\half}}F\left({\Sigma_1\over
\Sigma_2}\right)
\eqn\threeeleven
$$
where $F\left({\Sigma_1\over
\Sigma_2}\right)$ is a function symmetric under interchange of
$\Sigma_{1,2}$. In
the limit in which $\Sigma_2>>\Sigma_1$ the function $F$ behaves like
$\left(\Sigma_1\over \Sigma_2\right)^{\half}$. As in the 1+1 dimensional
case, the
energy gap is required to be surprisingly small as N increases.

The most interesting prediction about 2+1 \sym \ follows from the spatial
rotational invariance of string theory. The 2+1 dimensional \sym \ has explicit
$O(7)$ invariance associated with the 7 scalar fields. This invariance is
just the
explicit rotational invariance of the uncompactified transverse directions
of the
original M-Theory. The emergence of the new noncompact direction $Y$, in
the limit of vanishing torus size, requires the symmetry to be
enhanced to the group of 8 dimensional rotations. This of course only
happens in the strong coupling limit. For the abelian
$U(1)$ part of the theory the invariance can be made manifest. In addition
to the 7 scalar free fields, the bosonic content of the abelian theory
includes the  gauge field which in 2+1 dimensions can be dualized to give
an 8th free scalar field, completing a vector multiplet of $O(8)$. The
fermions also group into a spinor. From the point of view of this paper,
the $O(8)$ symmetry of the strongly coupled nonabelian theory
will be regarded as a prediction. Since in this case, as in the previous
example, strong coupling governs the infrared properties of the \sym, the
correct statement is that the fixed point describing the infrared has
$O(8)$ symmetry. Evidence for the correctness of the prediction has been
given in [\sethi,\IIAbs].

%
%
\chapter{The 3-torus}

Matrix theory compactified on a 3-torus leads to 3+1 dimensional $U(N)$
\sym \ with
4 supersymmetries. The compactification lengths are $L_1,L_2, L_3$ on the
M-theory
side and $\Sigma_1,\Sigma_2,\Sigma_3$ on the \sym \ side. The Yang-Mills
coupling
constant $g^2$ in this case is dimensionless. We find,
$$
g^2=\til =(2\pi)^4{l_{11}^3 \over L_1L_2L_3}
\eqn\fourone
$$
$$
\Sigma_i=(2\pi)^2{l_{11}^3 \over RL_i}
\eqn\fourtwo
$$
where $i=1,2,3$.
Now let us consider the states which become light as the $L_i$ shrink. The
integer
quantized magnetic fluxes can be labelled $n_{ij}$. We find the magnetic
energy is
given by
$$
E_M=(2\pi)^2\left({\Sigma_3 \over 2g^2 \Sigma_1 \Sigma_2}n_{12}^2 +   {\Sigma_1
\over 2g^2
\Sigma_2 \Sigma_3}n_{23}^2   +{\Sigma_2 \over 2g^2 \Sigma_3
\Sigma_1}n_{31}^2\right)
\eqn\fourthree
$$
For a single unit of flux in the 1,2 plane the energy is
$$
E_M=(2\pi)^2{\Sigma_3 \over 2g^2 \Sigma_1 \Sigma_2}= {R\over 2N}{L_1^2L_2^2
\over (2\pi)^4l_{11}^6}
\eqn\fourfour
$$
corresponding to a mass
$$
M_M={L_1L_2 \over (2\pi)^2l_{11}^3}
\eqn\fourfive
$$
and similarly for the other two  directions.
We interpret these light states as the Kaluza-Klein excitations of  3
decompactifying directions whose lengths are
$$
\bar L_3= (2\pi)^3{l_{11}^3 \over L_1L_2}
\eqn\foursix
$$
and so forth.

Evidently as the 3-torus shrinks, 3 new directions open up and restore the
theory
to an 11 dimensional theory. The only reasonable candidate is M-Theory
itself. We
will shortly see what the interpretation of this is in the \sym.

To determine the parameters of the new 11 dimensional theory we consider
the electric
energy for a field in the 1 direction.
$$
E_E ={g^2 \Sigma_1 \over 2N \Sigma_2 \Sigma_3}= {R\over 2N}{(2\pi)^2\over
L_1^2}
\eqn\fourseven
$$
leading to a  mass
$$
M_E={2\pi\over L_1}
\eqn\foureight
$$
The natural interpretation of this energy in the new M-Theory is a
membrane  wrapped around the 2,3 face of a growing 3 torus. The area of the
membrane is $\bar L_2\bar L_3$ and its energy is expected to have the form
${\bar L_2\bar L_3}\over (2\pi)^2{\bar l_{11}^3}$. Using eq. \foureight \ we
can identify
the new 11 dimensional planck length.
$$
\bar l_{11}^3= (2\pi)^3{l_{11}^6 \over L_1L_2L_3}
\eqn\fournine
$$
The new M-Theory must also be equivalent to a \sym \ whose parameters are
given in
terms of the barred versions of eqs. \fourone,\fourtwo . One finds
$$
\bar g = {2 \pi \over g}
\eqn\fourten
$$
This relation between coupling constants is just the S-duality of four
dimensional
\sym \ which interchanges electric and magnetic fields. In order that the new
11
dimensional theory be exactly M-Theory, the full $U(N)$ theory must have
electric-magnetic duality. In this way, if we did not already know of the
duality
of \sym, we might be led to predict it.

The full duality group of 3-dimensionally compactified M-theory is now easily
understood as the product of the $SL(2;Z)$ electric-magnetic duality and the
$SL(3;Z)$ geometric symmetry group of the 3-torus.

Thus far we have made use of the properties of 2- branes but we have not had
need for the 5-branes of M-Theory. Later we will find them indispensible.
However this is a good place to discuss them. The 5-branes have proved much
more elusive than the 2-branes which are found as classical configurations of
M(atrix) theory [\bss].  Following [\grt] we can gain some understanding of
5-brane
states by considering a 5-brane wrapped around the 3-torus. The other 2
dimensions  of the brane  form an unwrapped infinitely extended 2-brane.
This object together with an ordinary unwrapped 2-brane form a doublet
under the $SL(2;Z)$ duality. This can be seen by observing that under
T-duality the two configurations are interchanged. Since the T-duality
operation is represented by the electric-magnetic S-duality of \sym \ we
find that to construct the 5-brane it is necessary to understand the action
of S-duality on the fields of the \sym. Since electric-magnetic duality is
not a symmetry of  classical field theory, we do not expect the 5-brane to
be represented by a classical configuration. In fact  it probably can not
be represented locally in terms ot the fields of the original theory.   We
will however assume that it exists as a quantum state of
\sym. A
deeper understanding of the 5-brane will probably await a more concrete
operator construction of electric-magnetic duality.
%
%
\chapter{The 4 Torus}

Up to this point we have been mainly reviewing some
well known connections. In
this section we will see some surprising departures
from the pattern of the
previous sections. What we have seen up to now is that
light states associated
with magnetic flux are  interpreted in terms of new
dimensions of space. Thus if
we compactify $K$ dimensions and let the torus shrink,
the pattern has been that we regain $K(K-1)/2$
large
dimensions and the theory becomes $11+{K(K-3)\over 2}$
dimensional. If this
pattern were to persist, compactification on a shrinking 4-torus
would lead
to a 13
dimensional theory! This would be quite surprising since
there are no known
formulations of supergravity above 11 dimensions. We shall
see in what follows
that the interpretation of the light states is entirely
different for tori of
dimension 4 or higher. The analysis that we will present of
this case is based on an important observation of Rozali [\rozali ]. Rozali
observed that the group of dualities for 4 compact toroidal dimensions  is
$SL(5;Z)$, which also happens to be the group of symmetries of a 5-torus.
Rozali suggested that the 4+1 dimensional \sym \ is actually equivalent to
a 5+1 dimensional theory of some kind which lives on a 5-torus.  We will
see substantial evidence for this in what follows.

The connection between M-theoretic  and field theoretic quantities is
$$
\Sigma_i=(2\pi)^2{l_{11}^3 \over RL_i}
\eqn\fiveone
$$
$$
\til ={g^2\over (\Sigma_1 \Sigma_2 \Sigma_3 \Sigma_4)^{1\over 4}}  =
(2\pi)^4{l_{11}^3 \over (L_1L_2L_3L_4)^{3\over4}}
\eqn\fivetwo
$$
The magnetic and electric energies are
$$
E_M ={R\over2N}{L_i^2L_j^2 \over (2\pi)^4l_{11}^6}
\eqn\fivethree
$$
$$
E_E ={R\over2N}{(2\pi)^2\over L_i^2}
\eqn\fivefour
$$
and the corresponding masses are
$$
M_M= {L_iL_j\over (2\pi)^2l_{11}^3}
\eqn\fivefive
$$
$$
M_E={2\pi\over L_i}
\eqn\fivesix
$$
In addition to the electric and magnetic fluxes, a new quantum
number appears
in the 4+1 \sym \ that did not occur in the lower dimensional examples.
Ordinary Yang-Mills instanton solutions occur as static solitons. The new
quantum number  is just the integer valued  instanton charge  $Q$,
$$
Q= {1\over 32 \pi^2}\int d^4 \sigma  F\wedge F
\eqn\fiveseven
$$
The energy associated
with
 instanton number $Q$ is
$$
E_I={(2\pi)^2Q \over g^2}
\eqn\fiveeight
$$
The energies of the electric and magnetic fluxes were generally
quadratic in
the integer quantum numbers. This was essential to their
interpretation as
infinite momentum frame energies. By contrast, the energy stored in
instanton
charge is linear in $Q$. To interpret this we note that the energy of a
``photon" in the \sym \ is also linear in its integer valued wave number.
This suggests that $Q$ should be identified with wave number in a new 5th
direction  in the quantum field theory. In other words, in the limit that
$g^2$ becomes large, the 4-dimensional \sym \ somehow develops a new
spatial
direction.  The size of this
dimension can be read off from
eq. \fiveeight \ and is given by
$$
\Sigma_5 ={g^2\over 2\pi} =(2\pi)^5{l_{11}^6 \over RL_1L_2L_3L_4}
\eqn\fivenine
$$
This does not mean that a new dimension opens up in the spacetime M-theory.
What it does mean is that the original field theory on the 4-torus becomes
some kind of quantum field theory with the symmetry and low lying level
density of a theory on a five torus. In fact as the $L_i$  shrink this
fifth dimension becomes the largest of the $\Sigma_i$.

The geometric symmetry of the 5-torus is $SL(5;Z)$ which includes the
discrete five dimesnional  rotations. Thus the states and operators
of the
theory will be classified in $SL(5;Z)$ multiplets.
The first such multiplet
is the 4 field-theoretic momenta and the instanton
charge ${Q\over g^2}$. The
6 magnetic fluxes also must belong to a multiplet.
They can be combined with
the 4 components of electric field to form a 10
component antisymmetric
tensor. Let us denote the fluxes in the following way.
$$
\eqalign{\Phi_{ij}=&\epsilon_{ijkl}n_{kl} \cr
\Phi_{i5}=& m_i }
\eqn\fiveten
$$
where $n_{ij}$  and $m_i$ are  the integer valued magnetic and electric
fluxes.  We can now write the energies corresponding to a unit flux in the
symmetrical form
$$
E_{ab} ={\pi\over N}\left({\Sigma_a^2 \Sigma_b^2 \over
\Sigma_1\Sigma_2\Sigma_3\Sigma_4\Sigma_5}\right)
\eqn\fiveeleven
$$
where $a,b = 1,2,3,4,5$.

Let us now turn to the string theoretic interpretation
of the theory.  Consider the spectrum of string-like
excitations that can be
created by wrapping M-Theory 2-branes  around the cycles of
the 4-torus. There are
four obvious independent string states that correspond to a
membrane wrapped
around $L_1,L_2,L_3,L_4$. If the theory is to have the
symmetry of the 5-torus then there must be an additional 5th
state. In fact
there is such a string configuration arising from the wrapping of an
M-theory 5-brane around the 4-torus. In fact the formula for the
tensions of
the various types of strings exhibits a high degree of symmetry. Consider
first, the string tension of a 2-brane wrapped around $L_i$.
$$
T_i={L_i \over (2\pi)^2l_{11}^3}
\eqn\fivetwelve
$$
For the 5-brane wrapped on the 4-torus the tension is
$$
T_5={L_1L_2L_3L_4\over(2\pi)^5 l_{11}^6}
\eqn\fivethirteen
$$
These can be combined into a single formula
$$
T_a={1\over R \Sigma_a}
\eqn\fivefourteen
$$
which reveals the 5 dimensional symmetry.  From eq. \fivethirteen
we find that the string length scale for the wrapped 5-brane is given by
$$
{l_s^2\over 2}=\alpha'=(2\pi)^4\left ({l_{11}^6 \over L_1L_2L_3L_4} \right)
\eqn\fivefifteen
$$

It is now possible to understand the meaning of the light  magnetic
flux states
which occur as the M-theory  torus shrinks.  Assuming the four $L_i$
shrink in fixed proportion, the mass scale of these fluxes
tend to zero  as
$$
M^2\sim{L^4\over l_{11}^6}
\eqn\fivesixteen
$$
as seen from eq. \fivefive. Let us compare this with the tension of
the lightest of the various strings. In the limit $L\to 0$, the lightest
string is the wrapped 5 brane whose tension is given in  eq. \fivethirteen.
The two scales clearly agree with one another.  Evidently, the light
magnetic flux states are connected with the physics of the lightest strings.

To further understand the system as a string theory it is useful
to choose one of the four directions of the torus,  say $L_1$, and treat
it as the direction  identified with the dilaton. In other words we view
the system as type IIA string theory on a 3-torus with sides $L_2,L_3,L_4$.
We first permute the directions $L_2,L_3$.
Now we perform  T-duality, inverting the lengths $L_2,L_3$.   The identical
T-duality was studied in [ \tdual] where it was shown that the sides
$L_1,L_2,L_3$ transform into dual dimensions $\bar L_1,\bar L_2,\bar L_3,$
given by
$$
\eqalign{\bar L_1=& (2\pi)^3{l_{11}^3 \over L_2L_3} \cr
                \bar L_2=& (2\pi)^3{l_{11}^3 \over L_1L_3} \cr
            \bar L_3=&  (2\pi)^3{l_{11}^3 \over L_1L_2} \cr }
\eqn\fiveseventeen
$$
All the remaining dimensions are unaffected. Thus
$$
\bar L_4=L_4
\eqn\fiveeighteen
$$
In addition the T-dual theory has a transformed 11 dimensional Planck
 scale $\bar l_{11}$.
$$
\bar l_{11}^3 =(2\pi)^3{l_{11}^6 \over L_1L_2L_3}
\eqn\fivenineteen
$$

{}From eqs. \fivefifteen \  and \fiveseventeen \ it is seen that
the new
compactification scales are of the same order of magnitude
as the scale
governing the lightest strings, the wrapped 5-branes.
It is now clear what
the fate of the light magnetic fluxes is. We have already
seen that their energy
scale is the same as the wrapped 5-brane string scale.
The natural
interpretation of the 6 magnetic fluxes is as winding
 number and momentum modes
of the light strings wound on the three cycles of the
$\bar L$ torus. To make this correspondence explicit, first uplift the
type IIA theory to 11-dimensional M-theory on four radii $R_1,R_2,R_3,R_4$ such
that $R_4$ corresponds to the short direction $\bar L_4$. Now reduce
M-theory again to type IIA theory by compactifying on $R_4$.
This type IIA theory is weakly coupled, since $R_4$ is small.

The string coupling constant for the light strings is
easily found to be
$$
\bar g_s^2 ={L_1L_2L_3L_4^3 \over (2\pi)^3l_{11}^6}
\eqn\fivetwentyone
$$
which indicates that the string is weakly coupled.

In this picture, the light fluxes in the original picture become momentum
and winding modes. In particular, the correspondences are
$$
\eqalign{\phi_{23}=&momentum \ along \ 1 \cr
                \phi_{13}=&momentum \ along \ 2 \cr
                \phi_{12}=&momentum \ along \ 3 \cr
                \phi_{14}=&winding \ along \ 1 \cr
                \phi_{24}=&winding \ along \ 2 \cr
                \phi_{34}=&winding \ along \ 3 \cr}
\eqn\fivetwenty
$$

The result of the T-duality transformation can also be thought of
 as a new M(atrix)-theory description, equivalent to, but different
from the
original.  In fact it is related to the original description by a rotation of
the
5-torus
$\Sigma_1\Sigma_2\Sigma_3\Sigma_4\Sigma_5$ which rotates by $\pi /2$ in both
the
$4,5$ and $2,3$ planes.

In addition to the multiplet of 5 string states, the
theory contains other BPS multiplets of 0,  2  and 3 branes.
We have found
that these branes form 10, 5 and 10 dimensional multiplets of the 5
dimensional
rotations.

To understand the meaning of these results let us consider fixing $g$ and
varying the dimensionless coupling $\til$ by varying the overall scale of the
$\Sigma_{1,2,3,4}$. According to eq. \fivenine \ the size of the fifth
dimension $\Sigma_5$ is the fixed constant $g$. In other words the theory
behaves like a five dimensional system with a fixed size for the 5th
dimension and a varying size for the 4-torus. When the 4-torus,
$\Sigma_{1,2,3,4}$,  is very large, the behavior of the system at that scale is
4 dimensional. On the other hand, when the four torus, $\Sigma_{1,2,3,4}$ is
much smaller than  $\Sigma_5$ the system behaves in a 1+1 dimensional
fashion. Between these limits all 5 directions are equally important and the
dynamics is 5 dimensional.

%
%
\chapter{The 5 Torus}

We next consider compactification on a 5-torus. The analysis will follow
the same lines as the previous case.  We begin by noting that the U-duality
group in this case is $SO(5,5;Z)$. This implies that  \sym \ on a 5-torus also
has this symmetry and that the states should transform covariantly under it.
Thus we
are led to conclude that \sym \ in 5+1 dimensions  also has such a duality
structure. As an example consider states carrying  a unit of electric or
magnetic flux. There are 10 magnetic and 5 electric fluxes. Since there is
no 15-dimensional representation of $SO(5,5;Z)$, we must combine these states
with one or more  additional   states. If we return to the 4+1 dimensional
case we recall that the strings originating from 2-branes formed a multiplet
with the string formed from a wrapped 5-brane. If we compactify an additional
dimension it is obvious that we should combine the fully
wrapped 5-brane with the
wrapped 2-branes in a single multiplet. Since the 2-brane states are
identified
with magnetic flux we must group these quantities with the wrapped 5-brane
charge. As we have remarked previously it is not known how to represent this
quantity in terms of the \sym \ quantum fields. Assuming it exists it
combines with the 15 fluxes to form a 16 dimensional spinor of $SO(5,5;Z)$.
For simplicity, in what follows we will write the formulas for the special
case in which all the $L_i$ are equal.

T-duality can be used to transform the theory. Choosing the 1 direction to be
identified with the string coupling, we apply T-duality to the other 4
cycles.
$$
\eqalign{{\bar L_1 \over \bar l_{11}}=&(2\pi)^4{l_{11}^3 \over L^3} \cr
          {\bar L_{i} \over \bar l_{11}}=&2\pi \ \ (i=2,3,4,5) \cr
           \bar l_{11} = &(2\pi)^2{l_{11}^3 \over L^2} }
\eqn\sixone
$$

Thus in the dual theory, one of the compactification scales, $\bar L_1$
decompactifies while the other 4 are of order the 11 dimensional Planck
scale. This means that an element of the duality group has led us back to
M-theory, compactified on a 4-torus with radius of order unity in 11
dimensional Planck units. Another equivalent description is IIA string
theory on a 3-torus with  string coupling of order 1 and compactification
radius of order $\sqrt{\alpha'}$. Surprisingly, the corresponding gauge
theory  is 4+1 dimensional \sym \ at a value of $\til$ of order 1. Evidently,
this is the theory which governs the short distance behavior of 5+1
dimensional \sym.

In each of the above cases, the shrunken torus limit led to a theory which
was to some extent analyzable. In the 1-torus and 4-torus cases we found
weakly coupled string theories with compactification radii (in the 4-torus
case) of order the string scale. In the 3-torus case we were led back to 11
dimensional M-theory. For the 2 and 5 torus we found string theories at
intermediate coupling and no compactification radii smaller than $\sqrt
{\alpha'}$. In the case of the 6-torus things become worse. We are unable to
eliminate either infinite coupling or vanishing string compactification
radii. Either of these renders the theory unanalyzable by current methods. We
don't know what this means but it could potentially signal a
non-perturbative anomaly for toroidal compactification to  space-time
dimension lower than 6.

%
%
\chapter{Conclusions}
By combining the assumptions of M(atrix) Theory with known properties of
toroidally compactified M-theory we can derive properties of \sym \ , some of
which have not been previously known.  By considering the limit in which the
M-theory torus shrinks to zero size we obtain  information about the strongly
coupled fixed points which govern either the infrared or ultraviolet behavior
of the \sym.   These properties can either  be
thought of as predictions about \sym \ or as necessary conditions for M(atrix)
theory to correctly describe M-theory. The detailed predictions are listed in
the introduction, and will not be repeated here.

Many of the conclusions involve symmetries between different kinds of fluxes,
electric, magnetic and the mysterious 5-brane wrapping number. These fluxes
often play the role of momenta in various compact dimensions. Symmetries
between them frequently relate to space-time symmetries such as rotational
invariance [\sethi] as in the 2-torus case. There is one more quantity which is
not traditionally thought of as a flux but which also represents a discrete
momentum and that is $N$ itself. Indeed, the identification of $N/R$ as the
longitudinal momentum of the  infinite momentum frame was the
original basis for M(atrix) theory. This raises an interesting speculation.
Perhaps the entire collection of $U(N)$ theories are combined together into a
single ``master gauge theory"  in which $N$ appears as a flux. The
dualities of this master theory should contain operations which interchange
the usual fluxes with $N$. These dualities will be the basis for the full
spacetime symmetry group including the notoriously difficult symmetries which
rotate the longitudinal direction into transverse directions.

We already have some evidence that something like this is the case. Recall
that in section 2 we identified $D0$-branes as the quanized units of
energy  stored in the electric field. These $D0$-branes are not the original
\tild \ which carry longitudinal momentum but are related by the interchange
of a compact transverse direction with the longitudinal direction. The fact
that the parameters of the $D0$-branes come out correctly indicates that this
interchange is working  as it should.

All of this raises another issue which up to now we have swept under the rug.
Which of the results of the kind we have derived apply only in the large $N$
limit and which are valid for arbitrary $N$? From the viewpoint of [\bfss],
\sym \ only becomes M-theory in the $N\to \infty$ limit. We would like to
suggest that the validity of the M-theory, \sym \ connection is not really
limited in this way. In the literature on light cone quantization there is a
formulation called ``Discrete Light Cone Quantization" or DLCQ for
short[\DLCQ]. In DLCQ  the light like direction $X^-$ is assumed periodic with
radius
$R$ and the spectrum of the light-like momentum $p^{+}$ is discrete. The
starting point is slightly different than that in [\bfss] where a spacelike
direction
$X^{11}$ was taken to be compact. In the limit of large momentum $N \to
\infty$ the difference in not important but for finite momentum  it is. For
example in [\bfss] it was argued that systems with negative longitudinal
momentum such as anti-\tild s decouple  in the infinite momentum or large
$N$ limit.  In DLCQ the physical interpretation  also requires a large $N$
limit in which $N=p_{11} R$ tends to infinity. However in the DLCQ formulation
the negative $p_{11}$ systems decouple for all $N$, not just large N.

String theory can also be formulated in DLCQ. Basically it looks like
ordinary light cone quantization except for two things. The first is that all
longitudinal string  momenta   $p^+_{11}$ are discrete multiples of $1/R$. The
second difference is that strings which wind around the longitudinal direction
must be introduced. The suggestion we wish to make is that the matrix model at
finite  $N$ is the DLCQ formulation of M-theory.

Now one of the features of the DLCQ of string theory is that the T-dualities
associated with transverse directions is preserved even for finite $N$. The
only symmetries which are not operative at finite $N$ are those which mix
longitudinal and transverse directions. If this view is correct then the
various dualities studied in this paper should be valid at finite $N$. It is
interesting from this point of view that the electric magnetic duality of the
3-torus is expected to be correct at finite $N$. Furthermore, following
[\sethi] we can use this to prove that the rotational invariance is restored
as the 2-torus shrinks to zero with $N$ fixed. It then seems reasonable to
conjecture the validity of the $SL(5;Z)$ and $SO(5,5;Z)$ symmetries of the 4
and 5 torus.

\chapter{Acknowledgements}
We would like to thank Tom Banks and Barak Kol for discussions.

W. F. thanks the ITP at Stanford for its hospitality and was
supported in part by the Robert Welch Foundation and
NSF Grant PHY-9511632. A.R. was supported in part by
the Department of Energy
under contract no. DE-AC03-76SF00515. L.S. is supported
by NSF grant PHY-9219345.

\vfill
\refout
\end